\documentclass[12pt]{article}  
\usepackage{cite}
\usepackage{epsfig}
\usepackage{graphicx}
\usepackage{amsmath}
\usepackage{amssymb}
\usepackage{color}  

\usepackage{a41}
\usepackage{color}
\usepackage[rflt]{floatflt}              
\usepackage{float}
\usepackage{slashed}
\usepackage{epstopdf}

\def\asr{\left( \frac{\alpha_s}{4 \pi} \right)}
\def\b0{\beta_0}

\usepackage{array}

\usepackage[english]{babel}
\usepackage[latin1]{inputenc}
\usepackage[T1]{fontenc}
\usepackage{ae}

\usepackage{url}

\usepackage{amsmath, amsthm, amssymb}
\newtheorem{thm}{Theorem}[section]

\newtheorem{definition}[thm]{Definition}

\newcommand{\ep}{\varepsilon}

\usepackage{rotating}

\usepackage{graphicx}
\newcounter{mmacnt}
\def\restartmma{\setcounter{mmacnt}{0}}
\restartmma \catcode`|=\active
\def|#1|{\mathrm{#1}}
\catcode`|=12
\newenvironment{mma}{
 \par\smallskip
 \catcode`|=\active
 \parskip=0pt\parindent=0pt % locally
 \small
 \def\In##1\\{%
\def\linebreak{\hfill\break\null\qquad}%
\refstepcounter{mmacnt}
\hangindent=2.5em\hangafter=0
\leavevmode
\llap{\tiny\sffamily n[\arabic{mmacnt}]:=\kern.5em}%
\mathversion{bold}\footnotesize$\displaystyle##1$\normalsize
\mathversion{normal}\par
 }%
 \def\Print##1\\{%
\def\linebreak{\hfill\break}%
\hangindent=2.5em\hangafter=0
\leavevmode ##1\par}%
 \def\Out##1\\{%
\def\linebreak{$\hfill\break\null\hfill$}%
\kern\abovedisplayskip\par
\hangindent=2.5em\hangafter=0
\leavevmode
\llap{\tiny\sffamily Out[\arabic{mmacnt}]=\kern.5em}
\footnotesize$\displaystyle##1$\normalsize\hfill\null\par
\kern\belowdisplayskip
 }%
 \def\Warning##1##2\\{%
\def\linebreak{\hfill\break}%
\hangindent=2.5em\hangafter=0
\leavevmode
{\scriptsize##1 : ##2}\par}%
}{%
 \par\smallskip
}

\usepackage{color}

\newenvironment{fshaded}{%
\MakeFramed {\FrameRestore}
}%
{\endMakeFramed}

\allowdisplaybreaks[4]

\begin{document}
\setlength{\baselineskip}{0.515cm}
\sloppy
\thispagestyle{empty}
\begin{flushleft}
DESY 18--052 %\hfill {\tt arXiv:1804.xxxxx[hep-ph]}
\\
DO--TH 18/08
\end{flushleft}

\mbox{}
\vspace*{\fill}
\begin{center}

{\LARGE\bf Heavy Quark Form Factors at Three Loops} 

\vspace*{3mm}
{\LARGE\bf in the Planar Limit}

\vspace*{3mm} 

\vspace{3cm}
\large
{\large 
J.~Ablinger$^a$,
J.~Bl\"umlein$^b$, 
P.~Marquard$^b$,
N.~Rana$^b$
and 
C.~Schneider$^a$
}

\vspace{1.cm}
\normalsize
{\it $^a$~Research Institute for Symbolic Computation (RISC),\\
  Johannes Kepler University, Altenbergerstra{\ss}e 69,
  A--4040, Linz, Austria}

\vspace*{3mm}
{\it  $^b$ Deutsches Elektronen--Synchrotron, DESY,}\\
{\it  Platanenallee 6, D-15738 Zeuthen, Germany}

%%\today

\end{center}
\normalsize
\vspace{\fill}
\begin{abstract}
\noindent
 We compute the color-planar and complete light quark non--singlet contributions to the heavy quark form factors 
 in the case of the axialvector, scalar and pseudoscalar currents at three loops in perturbative QCD. 
 We evaluate the master integrals applying a new method based on differential equations for general
 bases, which is applicable for all first order factorizing systems. The analytic results are 
 expressed in terms of harmonic polylogarithms and real-valued cyclotomic harmonic polylogarithms.
\end{abstract}

\vspace*{\fill}
\noindent
% \numberwithin{equation}{section}
%%%%%%%%%%%%%%%%%%%%%%%%%%%%%%%%%%%%%%%%%%%%%%%%%%%%%%%%%%%%%%%%%%%%%%%%%%%%%%%%%%%%%%%%%%%%%%%%%%%%%%%%%%%%%%%%%%%%%%%%%%%%%%%%%%%
\newpage 

\noindent
Form factors are the matrix elements of local composite operators between physical states. 
In perturbative Quantum Chromodynamics (QCD) these objects play a significant role in determining 
physical observables. In scattering cross-sections, they provide important contributions to the 
virtual corrections. The massive form factors are of importance for the forward-backward asymmetry of 
bottom or top quark pair production at electron-positron colliders and to static quantities like the 
anomalous magnetic moment of a heavy quark and other processes. They are also of importance to scrutinize the 
properties of the top quark \cite{Abe:1995hr,D0:1995jca} during the high luminosity phase of the LHC \cite{HLHC} 
and the experimental precision studies at future high energy $e^+ e^-$ colliders \cite{Accomando:1997wt}.

In this letter, we calculate both the color--planar and complete light quark non-singlet three-loop 
contributions to the
massive form factors for axialvector, scalar and pseudoscalar currents. Our results for the vector current,
including a detailed account of the techniques used in these calculations, will be presented elsewhere
\cite{FORMF2}. The two-loop QCD corrections to  the massive vector, axialvector form factors, the anomaly 
contributions, and the scalar and pseudoscalar form factors were first presented in 
\cite{Bernreuther:2004ih,Bernreuther:2004th,Bernreuther:2005rw,Bernreuther:2005gw}. In \cite{Gluza:2009yy}, an 
independent computation led to a cross-check of the vector form factor, giving also  the additional 
${\mathcal O}(\ep)$ terms in the dimensional parameter $\ep = (4-D)/2$. Recently, the contributions up to 
${\mathcal O}(\ep^2)$ for all the massive two-loop form factors were obtained in Ref.~\cite{Ablinger:2017hst}.
The color--planar contributions to the massive three-loop form factor for the vector current have been computed 
in 
\cite{Henn:2016tyf,Henn:2016kjz} and the complete light quark contributions in \cite{Lee:2018nxa}.  
The large $\beta_0$ limit has been considered in \cite{Grozin:2017aty}.

Our notations follow those used in Ref.~\cite{Ablinger:2017hst}. We consider the decay of a virtual massive 
boson 
of momentum $q$ into a pair of heavy quarks of mass $m$, momenta $q_1$ and $q_2$ and color
$c$ and $d$, through a vertex $X_{cd}$, where
$X_{cd} = \Gamma^{\mu}_{A,cd}, \Gamma_{S,cd}$ and
$\Gamma_{P,cd}$ indicates the coupling to an axialvector, a
scalar and a pseudoscalar boson, respectively.  Here 
$q^2 = (q_1+q_2)^2$ is the center of mass
energy squared and the dimensionless variable $s$ is defined by
%-------------------------------------------------------------------------------------------------------------
\begin{equation}
 s = \frac{q^2}{m^2}\,.
\end{equation}
%-------------------------------------------------------------------------------------------------------------
The amplitudes take the following general form 
%-------------------------------------------------------------------------------------------------------------
\begin{equation}
 \bar{u}_c (q_1) X_{cd} v_d (q_2) \,,
\end{equation}
%-------------------------------------------------------------------------------------------------------------
where $\bar{u}_c (q_1)$ and $v_d (q_2)$ are the bi-spinors of the quark and the anti-quark, 
respectively.
We denote the corresponding UV renormalized form factors by $F_{I}$, with $I = A, S, P$.
They are expanded in the strong coupling constant $\alpha_s = g_s^2/(4\pi)$ as follows
%-------------------------------------------------------------------------------------------------------------
\begin{equation}
 F_{I} = \sum_{n=0}^{\infty} \asr^n F_{I}^{(n)} \,.
\end{equation}
%-------------------------------------------------------------------------------------------------------------
The following generic forms for the amplitudes can be found by studying the general Lorentz structure.
For the axialvector current, it can be cast as
%-------------------------------------------------------------------------------------------------------------
\begin{align}
 \Gamma_{A,cd}^{\mu}
 = -i \delta_{cd}\Big[
 a_Q \Big( \gamma^{\mu} \gamma_5~F_{A,1} 
         + \frac{1}{2 m} q^{\mu} \gamma_5 ~ F_{A,2}  \Big) \Big], 
\end{align}
%-------------------------------------------------------------------------------------------------------------
where $\sigma^{\mu\nu} = \frac{i}{2} [\gamma^{\mu},\gamma^{\nu}]$ and $a_Q$ is the 
Standard Model (SM) axialvector coupling constant.
Likewise, for the scalar and pseudoscalar currents, one has
%-------------------------------------------------------------------------------------------------------------
\begin{align}
 \Gamma_{cd} = \Gamma_{S,cd} +  \Gamma_{P,cd}
%  \nonumber\\
 = - \frac{m}{v} \delta_{cd} ~ \Big[ s_Q \, F_{S} + i p_Q \gamma_5 \, F_{P} \Big] \,,
\end{align}
%-------------------------------------------------------------------------------------------------------------
where $v = (\sqrt{2} G_F)^{-1/2}$ is
the SM vacuum expectation value of the Higgs field, with $G_F$ being the Fermi
constant, $s_Q$ and $p_Q$ are the scalar and pseudoscalar couplings, respectively.
Finally, to obtain the unrenormalized form factors, we multiply appropriate projectors
as provided in \cite{Ablinger:2017hst} and perform the trace over the color and spinor indices.
For later purpose we denote by $N_c$ the number of colors, and 
$n_l$ and $n_h$ are the number of light and heavy quarks, respectively. We 
will set $n_h = 1$ in the following.

Since we use dimensional regularization \cite{tHooft:1972tcz},
one important point is to define a proper description for the treatment of $\gamma_5$. 
Both the color-planar and complete $n_l$ contribution belong to the so-called non-singlet case, 
where the axialvector or pseudoscalar vertex is connected to open heavy fermion lines.
Hence, both $\gamma_5$-matrices appear in the same chain of Dirac matrices, which allows us
to use an anti-commuting $\gamma_5$ in $D$ space-time dimensions, with $\gamma_5^2 = 1$. This is implied 
by the well-known Ward identity, 
%-------------------------------------------------------------------------------------------------------------
\begin{equation} \label{eq:cwi}
 q^{\mu} \Gamma_{A,cd}^{\mu, \sf ns} = 2 m \Gamma_{P,cd}^{\sf ns} \,,
\end{equation}
%-------------------------------------------------------------------------------------------------------------
which in terms of the form factors, takes the  form
%-------------------------------------------------------------------------------------------------------------
\begin{equation} \label{eq:cwiFF}
 2 F_{A,1}^{\sf ns} + \frac{s}{2} F_{A,2}^{\sf ns} = 2 F_{P}^{\sf ns} \,.
\end{equation}
%-------------------------------------------------------------------------------------------------------------
Here ${\sf ns}$ denotes the non-singlet contributions. For convenience, we introduce 
the kinematic variable \cite{Barbieri:1972as}
%-------------------------------------------------------------------------------------------------------------
\begin{equation} \label{eq:varxp}
 x=\frac{\sqrt{q^2-4m^2}-\sqrt{q^2}}{\sqrt{q^2-4m^2}+\sqrt{q^2}}\quad \leftrightarrow 
\quad s = \frac{q^2}{m^2}=-\frac{(1-x)^2}{x},
\end{equation}
%------------------------------------------------------------------------------------------------------------
which we use in the following. In particular, we focus on the Euclidean region, $q^2<0$, corresponding to $x \in 
[0,1[$. 

%%%%%%%%%%%%%%%%%%%%%%%%%%%%%%%%%%%%%%%%%%%%%%%%%%%%%%%%%%%%%%%%%%%%%%%%%%%%%%%%%%%%%%%%%%%%%%%%%%%%%%%%%%%%%%%%%%%
\begin{figure}[ht]
\begin{center}
\begin{minipage}[c]{0.09\linewidth}
     \includegraphics[width=1\textwidth]{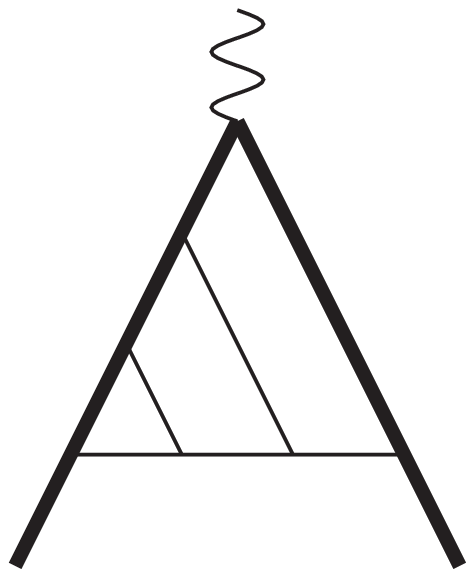}
\vspace*{-11mm}
\begin{center}
% {\footnotesize (a)}
\end{center}
\end{minipage}
\hspace*{2mm}
\begin{minipage}[c]{0.09\linewidth}
     \includegraphics[width=1\textwidth]{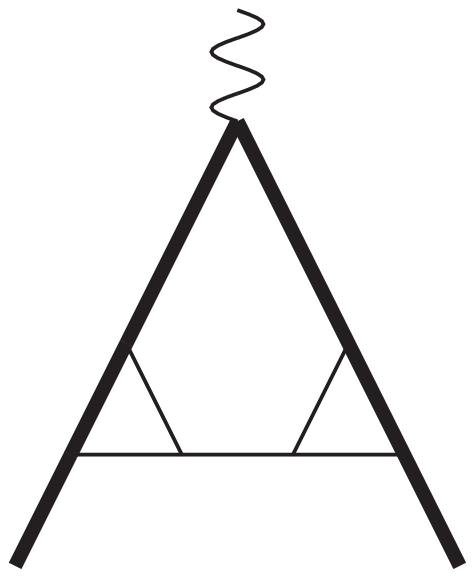}
\vspace*{-11mm}
\begin{center}
% {\footnotesize (b)}
\end{center}
\end{minipage}
\hspace*{2mm}
\begin{minipage}[c]{0.09\linewidth}
     \includegraphics[width=1\textwidth]{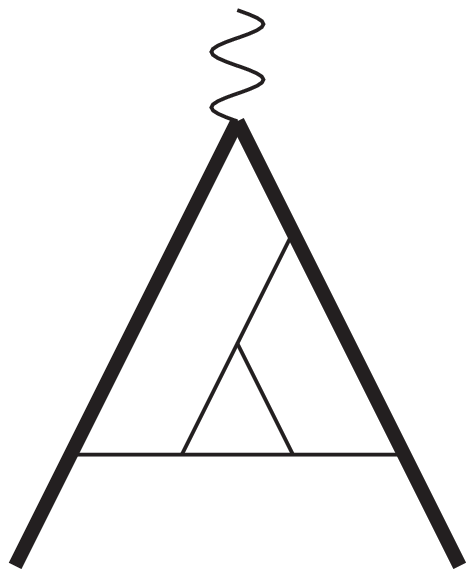}
\vspace*{-11mm}
\begin{center}
% {\footnotesize (c)}
\end{center}
\end{minipage}
\hspace*{2mm}
\begin{minipage}[c]{0.09\linewidth}
     \includegraphics[width=1\textwidth]{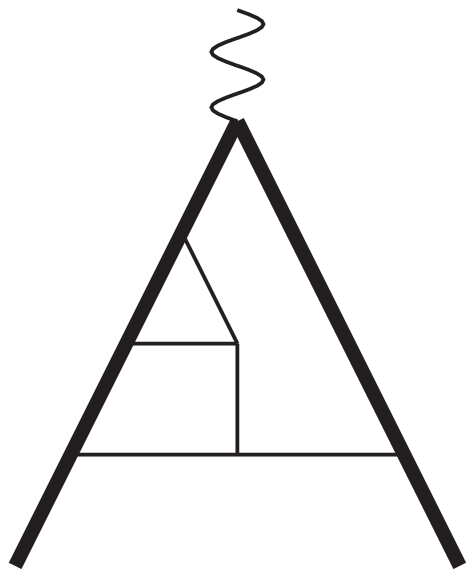}
\vspace*{-11mm}
\begin{center}
% {\footnotesize (a)}
\end{center}
\end{minipage}
\hspace*{2mm}
\begin{minipage}[c]{0.09\linewidth}
     \includegraphics[width=1\textwidth]{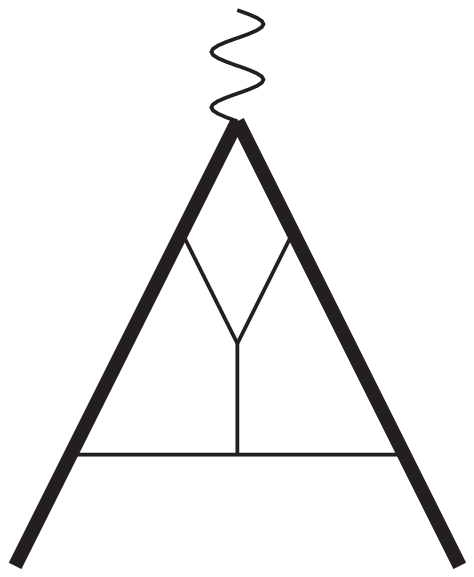}
\vspace*{-11mm}
\begin{center}
% {\footnotesize (b)}
\end{center}
\end{minipage}
\hspace*{2mm}
\begin{minipage}[c]{0.09\linewidth}
     \includegraphics[width=1\textwidth]{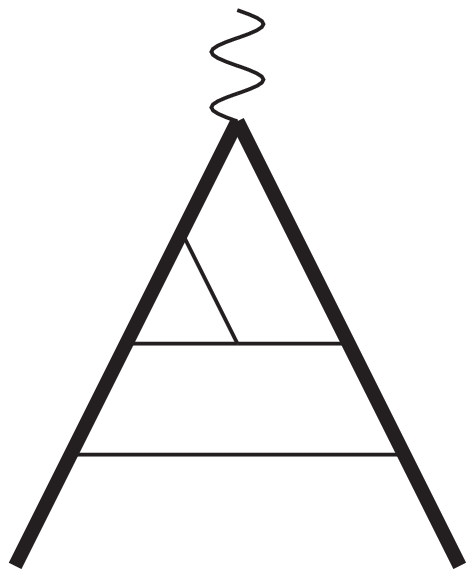}
\vspace*{-11mm}
\begin{center}
% {\footnotesize (b)}
\end{center}
\end{minipage}
\hspace*{2mm}
\begin{minipage}[c]{0.09\linewidth}
     \includegraphics[width=1\textwidth]{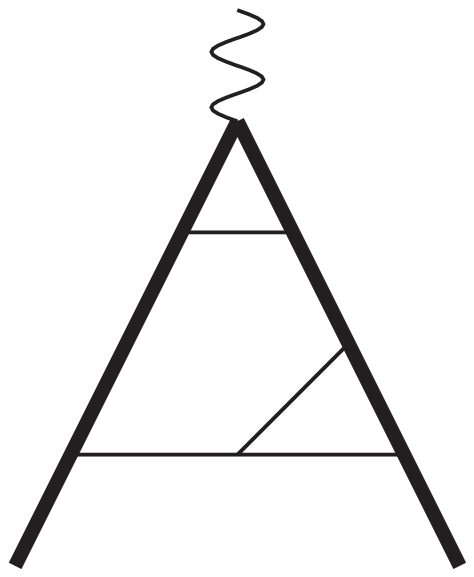}
\vspace*{-11mm}
\begin{center}
% {\footnotesize (b)}
\end{center}
\end{minipage}
\hspace*{2mm}
\begin{minipage}[c]{0.09\linewidth}
     \includegraphics[width=1\textwidth]{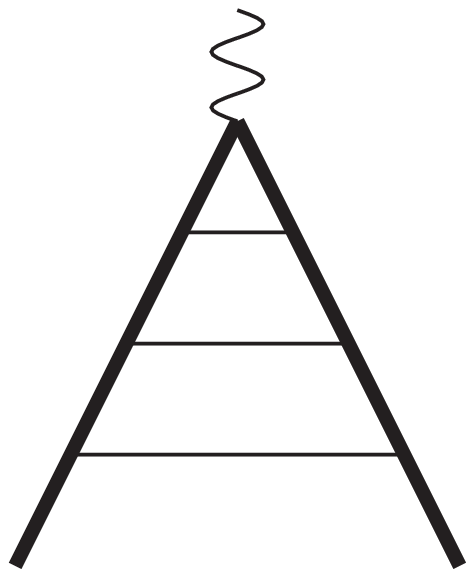}
\vspace*{-11mm}
\begin{center}
% {\footnotesize (b)}
\end{center}
\end{minipage}
\end{center}
\caption{\sf \small The color-planar topologies}
\label{fig:cptopologies}
\end{figure}
%%%%%%%%%%%%%%%%%%%%%%%%%%%%%%%%%%%%%%%%%%%%%%%%%%%%%%%%%%%%%%%%%%%%%%%%%%%%%%%%%%%%%%%%%%%%%%%%%%%%%%%%%%%%%%%%%%%
% 
\begin{figure}[H]
\begin{center}
    \begin{minipage}[c]{0.13\linewidth}
    \includegraphics[width=1\textwidth]{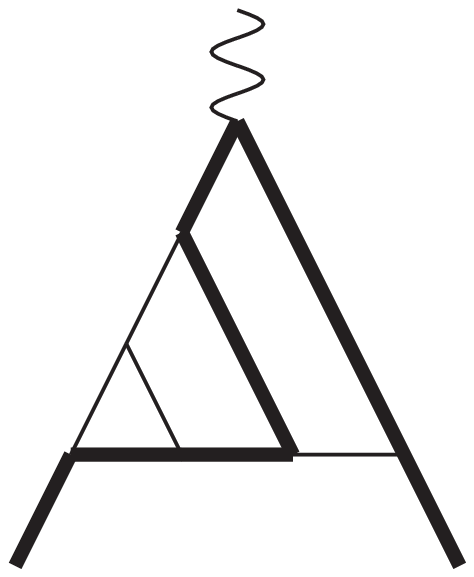}
    \vspace*{-11mm}
    \end{minipage}
        \hspace*{2mm}
    \begin{minipage}[c]{0.13\linewidth}
    \includegraphics[width=1\textwidth]{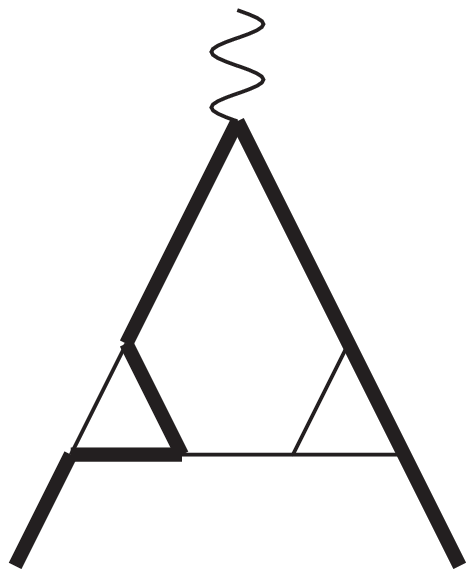}
    \vspace*{-11mm}
    \end{minipage}
        \hspace*{2mm}
    \begin{minipage}[c]{0.13\linewidth}
    \includegraphics[width=1\textwidth]{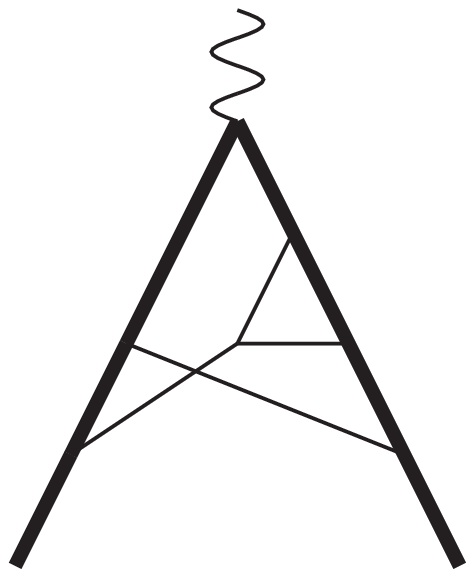}
    \vspace*{-11mm}
    \end{minipage}
\end{center}
\caption{\sf \small The $n_l$ topologies}
\label{fig:nltopo}
\end{figure}
%%%%%%%%%%%%%%%%%%%%%%%%%%%%%%%%%%%%%%%%%%%%%%%%%%%%%%%%%%%%%%%%%%%%%%%%%%%%%%%%%%%%%%%%%%%%%%%%%%%%%%%%%%%%%%%%%%%
% 
The Feynman diagrams for the different form factors are generated using {\tt QGRAF}
\cite{Nogueira:1991ex}, the color algebra is performed using {\tt Color} \cite{vanRitbergen:1998pn}, the output 
of which is then processed using {\tt Q2e/Exp} \cite{Harlander:1997zb,Seidensticker:1999bb} and {\tt FORM} 
\cite{Vermaseren:2000nd, Tentyukov:2007mu} in order to express the diagrams in terms of a linear combination of 
a large set of scalar integrals.  These integrals are then reduced using integration by parts identities (IBPs)
\cite{Chetyrkin:1981qh,Laporta:2001dd} with the help of the program {\tt Crusher} \cite{CRUSHER} to obtain
109 master integrals (MIs), out of which 96 appear in the color-planar case. 
In the color-planar limit, the families of integrals can be represented by eight topologies, shown in
Figure~\ref{fig:cptopologies}, whereas for the complete light quark contributions, three more topologies, 
cf.~Figure~\ref{fig:nltopo}, are required \footnote{Only sub-topologies with a maximum of eight propagators 
contribute.}.

Finally, the master integrals have to be computed. For this we use the method of differential equations, see 
also \cite{Kotikov:1990kg,Remiddi:1997ny,Henn:2013pwa,Ablinger:2015tua}. The corresponding differential equations
are obtained from the IBP relations. Here a central question is whether
the corresponding linear system of differential equations is first order factorizable or not. Using the 
package {\tt Oresys} \cite{ORESYS}, based on Z\"urcher's algorithm \cite{Zuercher:94,NewUncouplingMethod}, 
we have proved that the present system is indeed first order factorizable in $x$-space. Without any need to 
choose 
a special basis, one  is therefore in the position to solve the system in terms of iterated integrals of 
whatsoever alphabet, cf.~Ref.~\cite{FORMF2} for details. The differential equations are solved order by order
in $\varepsilon$ successively, starting at the leading pole terms $\propto 1/\varepsilon^3$. The successive 
solutions in $\varepsilon$ contribute to the inhomogeneities in the next order. We compute 
the master integrals block-by-block, where for an $m \times m$ system a single inhomogeneous ordinary 
differential equation of order $m$ or less is obtained, which we solved using the variation of constant.
The other $m-1$ solutions result from the former solution immediately. The boundary conditions can be 
determined 
by a separate calculation at $x = 1$. The calculation is performed by intense use of {\tt HarmonicSums}
\cite{HSUM,
Ablinger:2014rba, Ablinger:2010kw, Ablinger:2013hcp, Ablinger:2011te, Ablinger:2013cf, Ablinger:2014bra}, 
which uses the package {\tt Sigma} 
\cite{Schneider:sigma1,Schneider:sigma2}. We finally have checked all master 
integrals numerically using {\tt FIESTA} \cite{Smirnov:2008py, Smirnov:2009pb, Smirnov:2015mct}.

In the present case, the emerging harmonic polylogarithms stem from the inhomogeneities, adding further letters
which result from the rational coefficients in the differential equations. They are obtained by partial 
fractioning as the $k$-th powers of letters, $k \in \mathbb{N}$, which have to be transformed to 
the letters by partial integration in case. This method has some relation to the method of 
hyperlogarithms~\cite{Brown:2008um,Ablinger:2014yaa}. One obtains up to weight {\sf w=6} 
real-valued iterated integrals over the alphabet
%------------------------------------------------------------------------------------------------------------------
\begin{eqnarray}
\frac{1}{x},~~
\frac{1}{1-x},~~
\frac{1}{1+x},~~
\frac{1}{1-x+x^2},~~
\frac{x}{1-x+x^2},
\end{eqnarray}
%------------------------------------------------------------------------------------------------------------------
i.e.~the usual harmonic polylogarithms (HPLs) \cite{Remiddi:1999ew} and their cyclotomic extension 
\cite{Ablinger:2011te}, including the respective constants in the limit $x \rightarrow 1$, i.e. the multiple 
zeta 
values (MZVs) \cite{Blumlein:2009cf} and the cyclotomic constants \cite{Ablinger:2011te,Ablinger:2017tqs}.
In case of the iterated integrals we apply the linear representation. For a numerical implementation 
the use of the shuffle algebra \cite{Blumlein:2003gb} implemented in {\tt HarmonicSums} 
reduces the number of functions accordingly. In the MZV and cyclotomic case there are proven reduction relations
to weight {\sf w = 12} \cite{Blumlein:2009cf} and {\sf w = 5} \cite{Ablinger:2017tqs}, respectively, which we 
have 
used. The 64 cyclotomic constants which appear up to {\sf w = 5} reduce to 18. At {\sf w = 6} 124 cyclotomic 
constants remain at the moment. Note that there are more conjectured relations, cf.~\cite{Henn:2015sem}, based 
on PSLQ \cite{PSLQ}. If these conjectured relations are used, only multiple zeta values 
remain as constants in all form factors using our real representation for the cyclotomic harmonic polylogarithms.
The analytic result for the different form factors in terms of HPLs and cyclotomic HPLs 
\cite{Remiddi:1999ew,Ablinger:2011te} can be analytically continued  outside $x~\in~[0,1[$ by using the mappings 
$x \rightarrow -x, x \rightarrow (1-x)/(1+x)$ on the expense of extending the cyclotomy class in cases needed. 

The UV renormalization of the form factors has been performed in a mixed scheme.
We renormalize the heavy quark mass and wave function in the on-shell (OS)
renormalization scheme, while the strong coupling constant is renormalized 
in the $\overline{\rm MS}$ scheme, which is given by setting the universal factor $S_\varepsilon = 
\exp(-\varepsilon (\gamma_E - \ln(4\pi))$ for each loop order to one at the end of the calculation. The 
required renormalization constants are well known and are denoted by 
$Z_{m, {\rm OS}}$ \cite{Broadhurst:1991fy, Melnikov:2000zc,Marquard:2007uj,
Marquard:2015qpa,Marquard:2016dcn}, 
$Z_{2,{\rm OS}}$ \cite{Broadhurst:1991fy, Melnikov:2000zc,Marquard:2007uj,Marquard:2018rwx} and 
$Z_{a_s}$ \cite{Tarasov:1980au,Larin:1993tp}
for the heavy quark mass, wave function and strong coupling constant, respectively. 
For all the cases, the renormalization of the heavy-quark wave function and the
strong coupling constant are multiplicative, while the renormalization of massive fermion 
lines has been taken care of by properly considering the counter terms.
For the scalar and pseudoscalar currents, the presence of the heavy quark mass in the Yukawa coupling
employs another overall mass renormalization constant.

The infrared (IR) singularities of the massive form factors can be factorized \cite{Becher:2009cu}
as
a multiplicative renormalization factor. Its structure is
constrained by the renormalization group equation (RGE), as follows,
%-------------------------------------------------------------------------------------------------------------
\begin{equation}
 F_{I} = Z (\mu) F_{I}^{\mathrm{fin}} (\mu)\, ,
\end{equation}
%-------------------------------------------------------------------------------------------------------------
where $F_{I}^{\mathrm{fin}}$ is finite as $\ep \rightarrow 0$. The RGE for $Z(\mu)$ reads
%-------------------------------------------------------------------------------------------------------------
\begin{equation} \label{eq:rgeZ}
 \frac{d}{d \ln \mu} \ln Z(\ep, x, m, \mu)  = - \Gamma (x,m,\mu) \,,
\end{equation}
%-------------------------------------------------------------------------------------------------------------
where $\Gamma$ is the corresponding cusp anomalous dimension, which is by now available up to three-loop 
order \cite{Grozin:2014hna,Grozin:2015kna}. Notice that $Z$ does not carry any information regarding the vertex. 
Both $Z$ and $\Gamma$ can be expanded in a perturbative series in $\alpha_s$ as follows
%-------------------------------------------------------------------------------------------------------------
\begin{equation}
 Z = \sum_{n=0}^{\infty} \asr^n Z^{(n)} \,, \qquad
 \Gamma = \sum_{n=0}^{\infty} \asr^{n+1} \Gamma_{n}
\end{equation}
%-------------------------------------------------------------------------------------------------------------
and one finds the following solution for Eq.~(\ref{eq:rgeZ})
%-------------------------------------------------------------------------------------------------------------
\begin{align} \label{eq:solnZ}
 Z &= 1 + \asr \Bigg[ \frac{\Gamma_0}{2 \ep} \Bigg] 
   + \asr^2 \Bigg[ \frac{1}{\ep^2} \Big( \frac{\Gamma_0^2}{8} - \frac{\beta_0 \Gamma_0}{4} \Big) + \frac{\Gamma_1}{4 \ep} \Bigg] 
   \nonumber\\
  &+ \asr^3 \bigg[ \frac{1}{\ep^3} \left( \frac{\Gamma_0^3}{48} - \frac{\beta_0 \Gamma_0^2}{8} + \frac{\beta_0^2 \Gamma_0}{6} \right)
                 + \frac{1}{\ep^2} \left( \frac{\Gamma_0 \Gamma_1}{8} - \frac{\beta_1 \Gamma_0}{6} \right) 
                 + \frac{1}{\ep} \left( \frac{\Gamma_2}{6} \right) \bigg]
   + {\cal O} (\alpha_s^4) \,.
\end{align}
%-------------------------------------------------------------------------------------------------------------
Eq.~(\ref{eq:solnZ}) correctly predicts the IR singularities for all massive form factors at the
three-loop level. 

We finally obtain the color--planar and the complete light quark non--singlet  ($n_l$) contributions for 
the  three-loop  massive 
form factors for the same currents as before. Since 
the expressions are very long, we provide them as supplemental material along with this 
publication only.

\begin{figure}[htb]
\centerline{%
\includegraphics[width=0.49\textwidth]{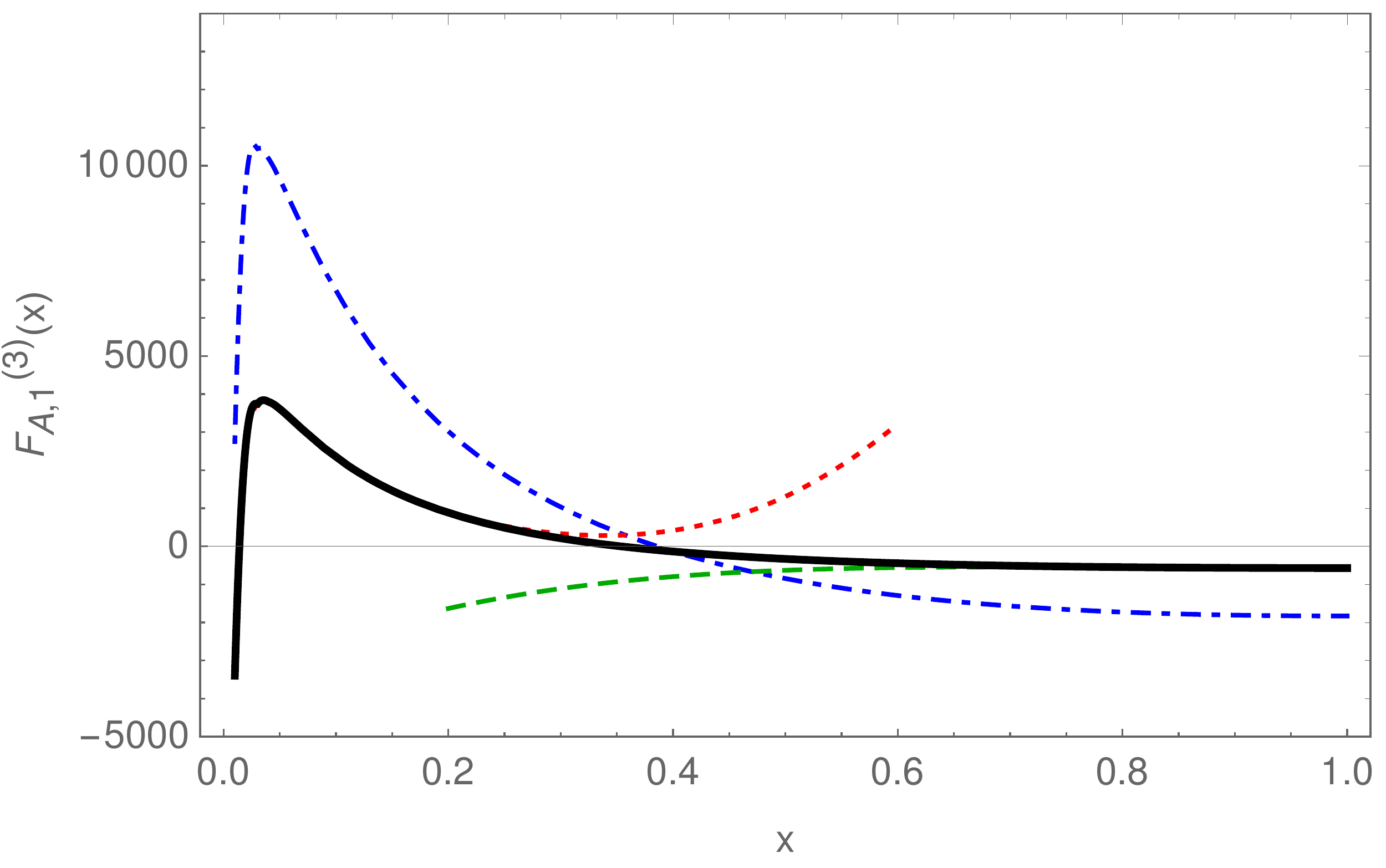}
\includegraphics[width=0.49\textwidth]{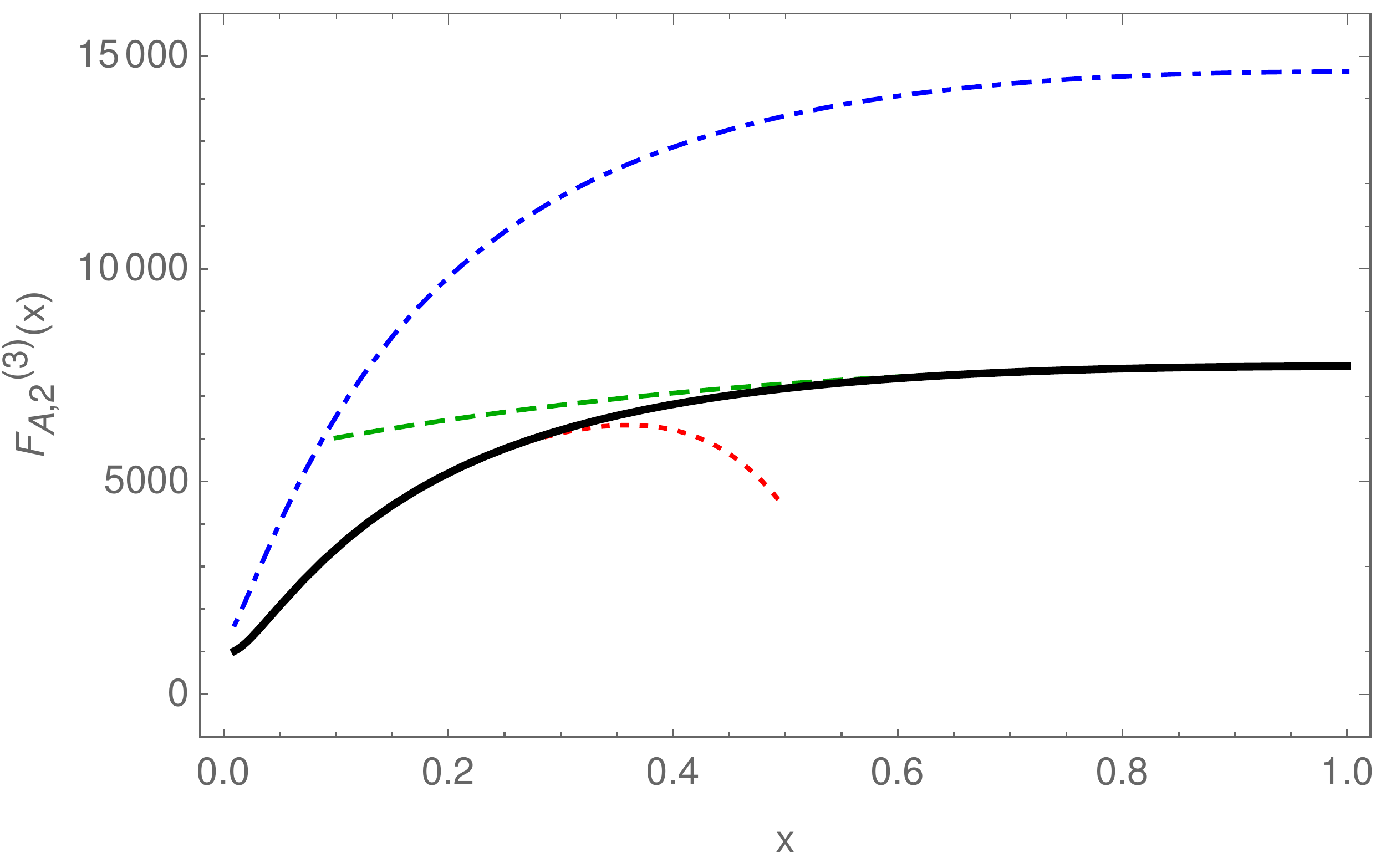}}
\caption{\sf The $O(\varepsilon^0)$ contribution to the axialvector three-loop form factors 
$F_{A,1}^{(3)}$ (left) and $F_{A,2}^{(3)}$ (right) as a function of $x$. 
Dash-dotted line: leading color 
contribution of the non-singlet form factor; Full line: sum of the complete non-singlet $n_l$-contributions for 
$n_l =5$ and the color-planar non-singlet form factor; Dashed line: large $x$ expansion; Dotted line: small $x$
expansion.}
\label{Fig:VF12cacfep1}
\end{figure}

\begin{figure}[htb]
\centerline{%
\includegraphics[width=0.49\textwidth]{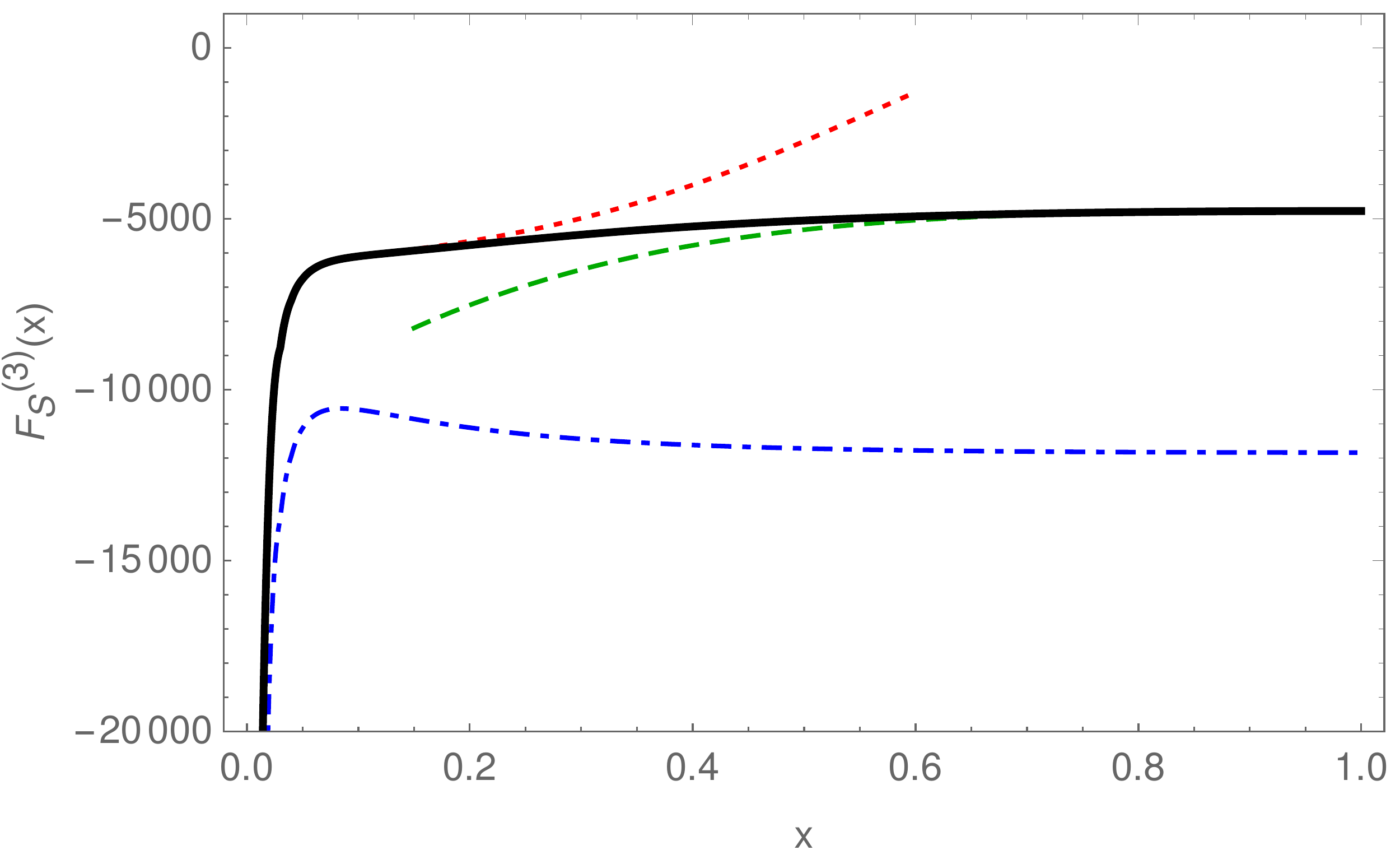}
\includegraphics[width=0.49\textwidth]{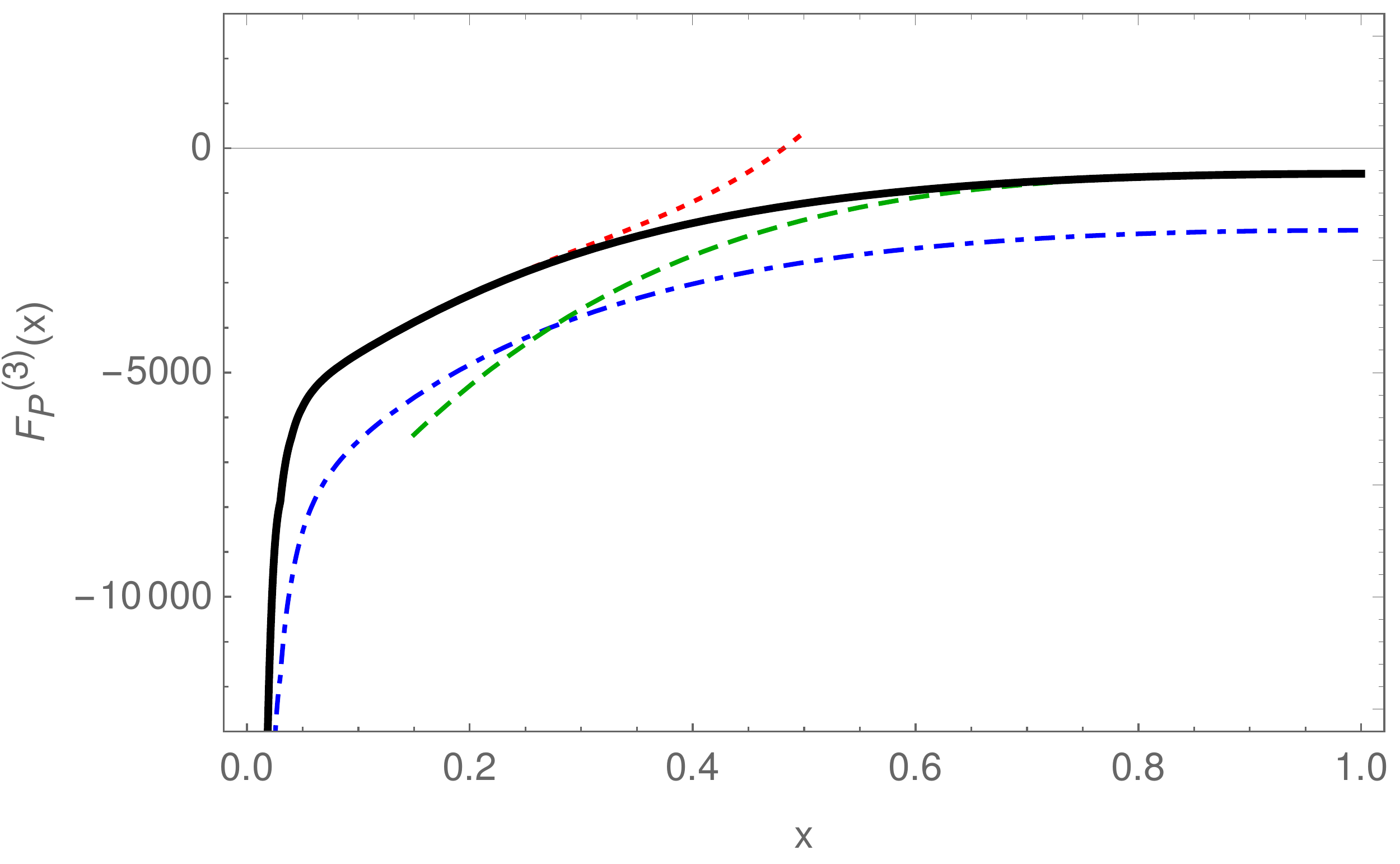}}
\caption{\sf The $O(\varepsilon^0)$ contribution to the scalar and pseudoscalar three-loop form factors 
$F_S^{(3)}$ (left) and $F_P^{(3)}$ (right) as a function of $x$. 
Dash-dotted line: leading color 
contribution of the non-singlet form factor; Full line: sum of the complete non-singlet $n_l$-contributions for 
$n_l =5$ and the color-planar non-singlet form factor; Dashed line: large $x$ expansion; Dotted line: small $x$
expansion.}
\label{Fig:AVG12cacfep1}
\end{figure}

In Figures~\ref{Fig:VF12cacfep1}--\ref{Fig:AVG12cacfep1} we illustrate the behaviour of the $O(\varepsilon^0)$ 
parts 
of the different form factors
as a function of $x \in [0,1]$. We also show their small- and large-$x$ expansions. The latter 
representations are obtained using {\tt HarmonicSums}. The different limits are characterized as follows :

\noindent 
\emph{Low energy region} ($x \rightarrow 1$): In the space-like case ($q^2 < 0$) we expanded the
form factors, redefining $x=e^{i\phi}$, $\phi=0$.

\noindent
\emph{High energy region} ($x \rightarrow 0$): 
Here we expand the form 
factors up to ${\cal O} (x^4)$. The chirality flipping form factors $F_{V,2}$ and 
$F_{A,2}$ vanish
and the effect of $\gamma_5$ gets nullified in this limit implying $F_{V,1}=F_{A,1}$ and $F_S=F_P$.

\noindent
\emph{Threshold region} ($x \rightarrow -1$): Here expansions of the form factors in $\beta = \sqrt{1 - 
\frac{4m^2}{q^2}}$ describe the dominant terms.

\noindent
For the numerical evaluation of the HPLs and the cyclotomic  HPLs we use the {\tt GiNaC}-package
\cite{Vollinga:2004sn,Bauer:2000cp}.

\noindent
We performed a series of further checks. The Ward identity (7) has been checked by an explicit calculation.
By maintaining the gauge parameter $\xi$ to first order, a partial 
check on gauge invariance has been obtained. After $\alpha_s$-decoupling the UV renormalized results satisfy the 
universal IR structure, confirming again the correctness of all pole terms. 
Finally, we compared our results with those of Ref.~\cite{LEE2018}, which has been obtained using partly 
different methods, and agree by adjusting the respective conventions.

\vspace{2ex}
\noindent
{\bf Acknowledgment.}~
This work was supported in part by the Austrian
Science Fund (FWF) grant SFB F50 (F5009-N15). We would like to thank M.~Steinhauser for providing their yet 
unpublished results in electronic form and A.~De Freitas and V.~Ravindran for discussions. The Feynman diagrams 
have been drawn using {\tt Axodraw} 
\cite{Vermaseren:1994je}.

\small
%-------------------------------------------------------------------------------------------------------
\providecommand{\href}[2]{#2}\begingroup\raggedright

%-------------------------------------------------------------------------------------------------------
\endgroup
%-------------------------------------------------------------------------------------------------------
\end{document}